\documentclass[traditabstract]{aa}
\usepackage{txfonts}
\usepackage{graphicx}
\usepackage{natbib}
\usepackage{amssymb}
\usepackage{paralist}

\newcommand{\binko}[2]{
\left(\!\begin{array}{c}{#1}\\{#2}\end{array}\!\right) }

\begin{document}

\title{Morphology of the Local Volume}
\author{
Martin Kerscher\inst{1}
\and
Anton Tikhonov\inst{2}
}

\institute{
Mathematisches Institut, Ludwig--Maximilians--Universit\"{a}t, Theresienstrasse 39, 
D--80333 M\"{u}nchen, Germany, \email{kerscher@math.lmu.de}
\and
Saint-Petersburg State University, Russian Federation, \email{avtikh@gmail.com}
}

\date{\today}

\abstract{To study the global morphology of the galaxy distribution in
our local neighbourhood we calculate the Minkowski functionals for a
sequence of volume limited samples within a sphere of 8\,Mpc centred
on our galaxy. The well known strong clustering of the galaxies and
the dominance of voids and coherent structures on larger scales is
clearly visible in the Minkowski functionals.
The morphology of the galaxy distribution changes with the limiting
absolute magnitude. The samples, encompassing the more luminous
galaxies, show emptier voids and more pronounced coherent
structures. Indeed there is a prominent peak in the luminosity
function of isolated galaxies for $M_B\approx-14$, which at least
partly explains these morphological changes.
We compare with halo samples from a $\Lambda$CDM simulation. Special
care was taken to reproduce the observed local neighbourhood as well
as the observed luminosity function in these mock samples.
All in all the mock samples render the global morphology of the galaxy
distribution quite well.  However the detailed morphological analysis
reveals that real galaxies cluster stronger, the observed voids are
emptier and the structures are more pronounced compared to the mock
samples from the $\Lambda$CDM simulation.}


\maketitle

%%%%%%%%%%%%%%%%%%%%%%%%%%%%%%%%%%%
\section{Introduction}

We study the global morphology of the galaxy distribution in the local
volume using Minkowski functionals and compare the geometry and
topology of the galaxy distribution to halo samples from a
$\Lambda$CDM-simulation. Only in our local neighbourhood we are able
to observe objects down to very low absolute luminosities and only
there we will be able to construct an almost complete galaxy sample.
\cite{karachentsev:catalog} presented the essential version of such a
catalogue of neighbouring galaxies. They outlined the main structural
elements of the local volume: the local sheet (part of the local
supercluster), the groups containing about 2/3 of the total local
volume population, and the local (Tully) void (and some other voids).
\cite{tikhonov:minivoids} and \cite{tikhonov:emptiness} analysed these
voids and the galaxy distribution within.
To study the global morphology of the galaxy distribution we use the
same data -- an updated catalogue of neighbouring galaxies
(Karachentsev, private communication).

Cold dark matter simulations together with a cosmological constant
($\Lambda$CDM-simulations) are considered to match the distribution of
$M^*$-like galaxies quite well. However on small scales there are
problems with the abundance of small mass aggregations. The
$\Lambda$CDM model predicts thousands of dwarf dark matter halos in
the local group \citep{klypin:missing, moore:dark, madau:dark}, while
only $\sim50$ are observed. Recently \cite{tikhonov:emptiness} found in
$\Lambda$CDM-simulations a severe overabundance (a factor of 10) of
halos \emph{in the voids} compared to the observed number of galaxies
in nearby voids.
The overabundance of halos on small mass scales as well as in the
voids is certainly a problem for the $\Lambda$CDM-model.  In the
present study, we investigate the spatial distribution of halos from a
$\Lambda$CDM simulation compared to the galaxies observed in the local
volume.
For the comparison with halo samples, we use the procedure from
\cite{tikhonov:emptiness} to map the circular velocities of the dark
matter halos to luminosities. This allows the construction of mock
galaxy samples with the same number of objects and the same luminosity
function as observed in the galaxy sample from the local volume.
Hence, we assume that the $\Lambda$CDM overabundance can either be
solved by the detection of new (very) low surface brightness galaxies,
or by mechanisms suppressing the galaxy formation in small dark halos
(see \citet{tikhonov:emptiness} for a discussion and references). Our
focus is on the global morphology of the galaxy distribution compared
to the halo distribution.

\subsection{The galaxy sample}
\begin{figure}
\resizebox{\hsize}{!}{\includegraphics{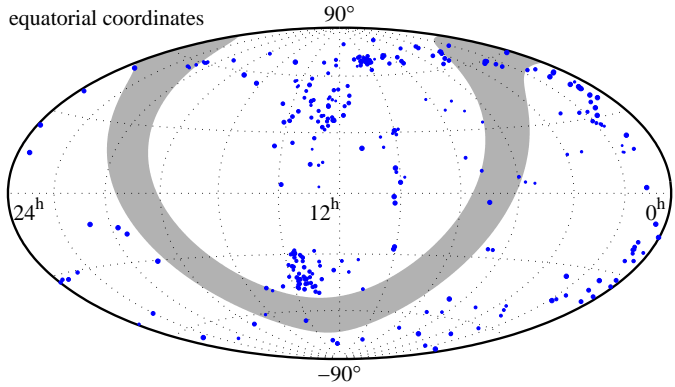}}
\caption{The distribution of the galaxies in the local volume on the sky.}
\label{fig:sky}
\end{figure}

Over the past few years searches for galaxies with distances less than
10\,Mpc have been undertaken using numerous observational data
including searches for low surface brightness galaxies, blind HI
surveys, and NIR and HI observations of galaxies in the zone of
avoidance \citep{karachentsev:catalog,karachentsev:mining}. The sample
contains about 550~galaxies (see Fig.\,\ref{fig:sky}),
the distances to the galaxies are measured independent
of the redshift, mostly with tip of the red giant brunch stars,
cepheids, the Tully-Fisher relation, and some other secondary distance
indicators.  The galaxies in the local volume sample map the ``real'' 3D
distribution.  The absence of the ``fingers of God'' effect simplifies
the morphological analysis of the local volume.
The distances have been measured with an accuracy of 8-10\% for most
of the galaxies \citep{karachentsev:catalog}.  According to Karachentsev
(2007, private communication) this sample is nearly volume limited
within an 8\,Mpc radius from our galaxy, and is reasonably complete for
galaxies with an absolute magnitudes $M_B\le-12$.
Some density gradients visible in the local volume sample are not
necessarily caused by an incompleteness. They can be explained more
naturally with structures in the local volume. The majority of groups
resides in a thin layer with a distance of $<0.3$\,Mpc from the
supergalactic plane.  The filaments of the local supercluster fall
mainly in the supergalactic plane, causing an excess of galaxies in
this region compared to the outer regions perpendicular to the
supergalactic plane.
The overall completeness of this sample is discussed in
\citet{tikhonov:emptiness}. They concluded that a significant
incompleteness of the sample is improbable. There still exists the
possibility that the observational sample is incomplete, missing very
low surface brightness galaxies free from gas, beyond the sensitivity
of modern ground-based telescopes.
The limiting magnitudes and the number of galaxies in the subsamples
considered are summarised in Table~\ref{tab:gal}.
\begin{table}
\caption{The galaxy samples considered.}\label{tab:gal}
\centering
\begin{tabular}{c|c|c}
\hline\hline
\textbf{sample} & \textbf{n.o.\ galaxies} & \textbf{magnitude cut}\\
\hline
\texttt{lv8m12} & 315 & $M_B\le-12$ \\
\texttt{lv8m14} & 205 & $M_B\le-14$ \\
\texttt{lv8m15} & 133 & $M_B\le-15$ \\
\texttt{lv8m16} & 96  & $M_B\le-16$ \\
\hline
\end{tabular}
\end{table}

%%%%%%%%%%%%%%%%
\subsection{Minkowski functionals}

\begin{figure}
\resizebox{\hsize}{!}{\includegraphics{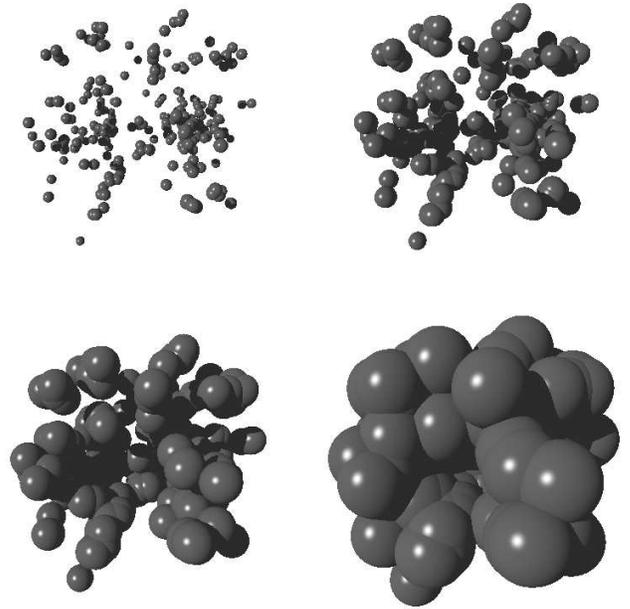}}
\caption{Galaxies from the local volume decorated with balls of
varying radii.
\label{fig:spheres}}
\end{figure}

The genus of an isodensity surface is a well known method to describe
the topology of cosmological density fields
{}\citep{gott:quantitative}. Minkowski functionals provide a unifying
framework for the topology and the geometrical quantities like volume,
surface area, and integrated mean curvature.  They have been developed
for the morphological characterization of the large scale distribution
of galaxies and galaxy clusters by \citet{mecke:robust} and have
successfully been used in cosmology
\citep{kerscher:abell,kerscher:statistical}, porous and disordered
media, dewetting phenomena, fluid and magneto hydrodynamics (see
\cite{mecke:additivity} for a review).

In order to quantify the spatial distribution of the galaxies we
decorate the galaxies with balls of varying radii (see
Fig.\,\ref{fig:spheres}).
Consider the union set $A_r=\bigcup_{i=0}^N B_r(\mathbf{x}_i)$ of
balls of radius $r$ around the $N$ galaxies at positions
$\mathbf{x}_i$, $i=1,2,\dots N$, thereby creating connections between
neighbouring balls.  The global morphology of the union set of these
balls changes with radius $r$, which is employed as a diagnostic
parameter.
It seems sensible to request that global geometrical and topological
valuations of e.g.\ $A_r$ are additive, invariant under rotations and
translations, and are continuous, at least for convex bodies. With
these prerequisites \citet{hadwiger:vorlesung} could show that in
three dimensions the four Minkowski functionals $M_{\mu}(A_r)$, $\mu =
0,1,2,3$, give a complete morphological characterization of the body
$A_r$.
The Minkowski functional $M_0(A_r)$ simply its volume, $M_1(A_r)$
is an eight of its surface area, $M_2(A_r)$ is its mean curvature
divided by $2\pi^2$, and $M_3(A_r)$ its Euler characteristics
multiplied by $\frac{3}{4\pi}$.
Volume and surface area are well known quantities.  The integral mean
curvature and the Euler characteristic are defined as surface
integrals over the mean and the Gaussian curvature respectively.  This
definition is only applicable for bodies with smooth boundaries.  In
our case we have additional contributions from the intersection lines
and intersection points of the balls.  \cite{mecke:robust} discuss the
extension for a union set of convex bodies.

One may express Minkowski functionals in terms of $n$--point
correlation functions, similar to a perturbative expansion.  But the
strength of an analysis with Minkowski functions is the direct
quantification of the morphology.
The additivity of the local contributions to the Minkowski functionals
ensures the robustness of the Minkowski functionals even for small
point sets \citep{mecke:robust}.  In an analysis with Minkowski
funtionals, using up to 385 galaxy clusters from the \textsc{Reflex}
sample, \cite{kerscher:reflex} were able to clearly distinguish the
distribution of observed galaxy clusters from a Gaussian point
distribution.  As can bee seen from the Table~\ref{tab:gal} we are
confronted with samples with a similar (small) number of points.
Closely related to our analysis is the investigation of the void
sizes, as performed by \cite{tikhonov:emptiness} for the same
catalogue of nearby galaxies.  Indeed the void probability
distribution function is equal to one minus the volume density, the
first Minkowski functional. Hence our analysis complements these
investigation by additionally using the other three Minkowski
functionals, to quantify the geometry, shape and topology of the
galaxy sample.

%%%%%%%%%%%%%%%%%%%%%%%%%%%%%%%%%%%
\section{Morphology of the local Volume}
\label{sec:morph-local-vol}

\begin{figure*}
\sidecaption
\includegraphics[width=12cm]{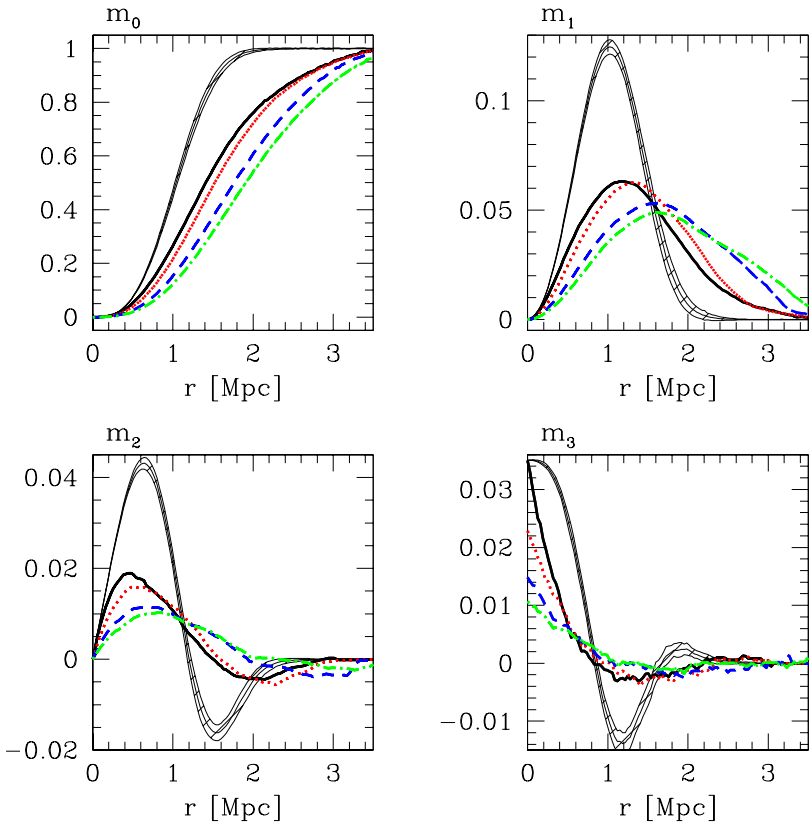}
\caption{The volume densities $m_\mu(A_r)$ of the Minkowski
  functionals determined from the galaxy samples \texttt{lv8m12}
  (black, solid), \texttt{lv8m14} (red, dotted), \texttt{lv8m15}
  (blue, dashed), \texttt{lv8m16} (green, dashed dotted) For reference
  the $m_\mu(A_r)$ of a Poisson process with the same number density
  as \texttt{lv8m12} are shown (one-$\sigma$ area, thin black lines).}
\label{fig:min_gal}
\end{figure*}

The galaxy and halo samples considered are only complete within a
sphere of 8\,Mpc radius. We use boundary corrections as discussed in
Appendix\,\ref{sec:calculating} to calculate the volume densities of the
Minkowski functionals in an unbiased way.
In Appendix\,\ref{sec:models} we show how Minkowski functionals
describe the morphological features and briefly discuss some
stochastic models.

In Fig.\,\ref{fig:min_gal} we show the volume densities $m_\mu(A_r)$
of the Minkowski functionals for a series of volume limited samples
from the local galaxy distribution. For reference the functionals of
randomly distributed points (Poisson process) are shown.\\
Compared to randomly distributed points, the volume density $m_0(A_r)$
increase is considerably delayed for all of the galaxy samples. The
empty space in between the strongly clustering galaxies fills up later
than for the purely random distribution.\\
The balls on the clustering galaxies already overlap for small radii,
causing a slow rise and a reduced maximum of the surface density
$m_1(A_r)\times8$ compared to Poisson distributed points. For large
radii the surface density of the galaxies is above the values for
randomly distributed points. The galaxies cluster on low dimensional
structures and consequently the balls have more room to grow.
In section~\ref{sec:morph-mag} we will comment on the excess surface
density in the samples with absolute magnitude $B\le-15$, as its is
visible for large radii.\\
Again, due to the clustering, the density of the integral mean
curvature $m_2(A_r)2\pi^2>0$ increases slower and the maximum is
reduced in comparison to a Poisson process.
In a Poisson process we get completely enclosed voids leading to the
strong negative signal. However from the galaxy samples only a week
negative signal is seen. The concave structures are less prominent, as
expected for clustering on planar (super galactic plane) or even
filamentary
structures.\\
For the radius $r=0$ the density of the Euler characteristic
$m_3(A_0)\frac{4\pi}{3}$ equals the number density of the galaxies in
the sample.  Hence the difference at $r=0$ is only a reflection of the
different sampling.  With increasing radius more an more balls overlap
and the Euler characteristic decreases. Then tunnels through the
structure are forming, giving a negative contributions to $m_3(A_r)$.
In the galaxy distribution only a small positive values for $m_3(A_r)$
can be observer, strengthening the previous observation that the voids
are not completely enclosed.

%%%
\subsection{Morphology changing with the absolute Magnitude}
\label{sec:morph-mag}

\begin{figure*}
\sidecaption
\includegraphics[width=12cm]{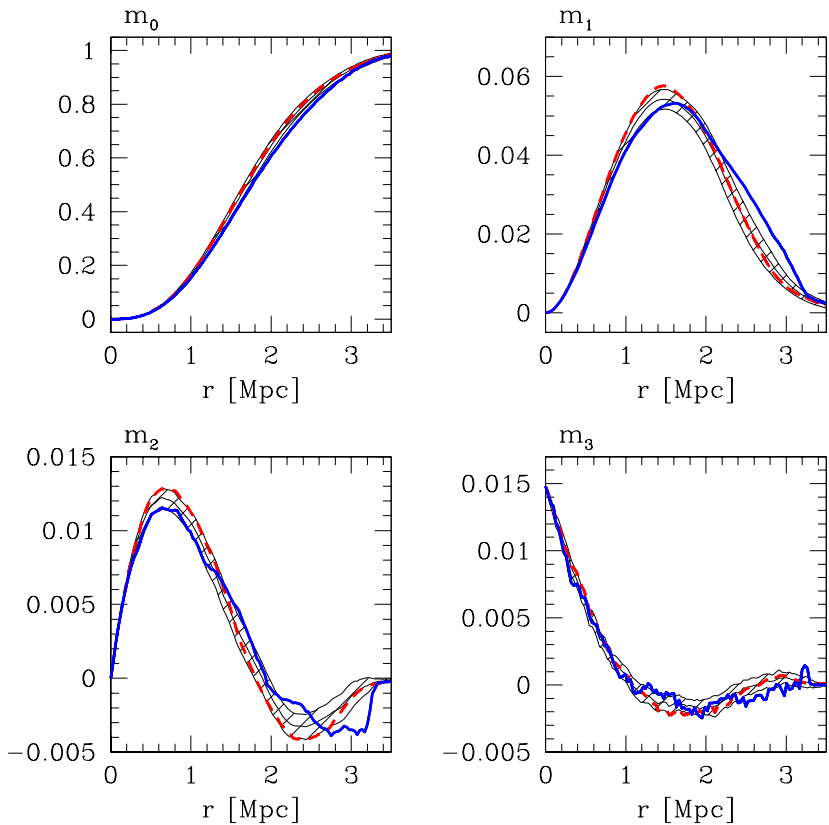}
\caption{Volume densities $m_\mu(A_r)$ of the Minkowski functionals
  determined from the galaxy samples \texttt{lv8m15} (blue solid) compared
  to the subsampled galaxy samples \texttt{lv8m14} (red dashed) and
  \texttt{lv8m12} (black, one-$\sigma$ error from randomly
  subsampling).}
\label{fig:minm_galm1415}
\end{figure*}

Already from Fig.\,\ref{fig:min_gal} we get the impression that the
morphology of the galaxy distribution changes significantly if we
include galaxies with absolute magnitudes $M_B\le-15$ in our
analysis. However we have to be more careful.
Minkowski functionals calculated from points decorated with balls do
depend on the number density of the point distribution. One can derive
explicit expression in terms of high order correlation functions
quantifying this non-trivial dependence on the number density (see
e.g.~\citealt{mecke:additivity}). 
To compare the galaxy distributions with different limiting magnitudes
we generate samples with the same number of points. We randomly
subsample the galaxy samples \texttt{lv8m12} and \texttt{lv8m14} to
the same number density as in the sample \texttt{lv8m15}.  This allows
us to compare the volume densities $m_\mu(A_r)$ for the galaxy sample
with $M_B\le-12$, $M_B\le-14$ and $M_B\le-15$ in
Fig~\ref{fig:minm_galm1415}.
The Minkowski functionals of the samples with $M_B\le-12$ and
$M_B\le-14$ agree within the error bars. Also the Volume density and
the Euler characteristic mostly agree between all the samples, whereas
the sample with $M_B\le-15$ shows an increased surface density
$m_1(A_r)$ for radii from 2.2 to 3.2\,Mpc. The voids become
significantly more emptier going from limiting magnitude $M_B\le-14$
to $M_B\le-15$. For the density of the integral mean curvature
$m_2(A_r)$ we see that the negative contributions occur only for
larger radii in the $M_B\le-15$ sample. These emptier voids are
surrounded by concave structures at larger radii.

\begin{figure}
\resizebox{\hsize}{!}{\includegraphics{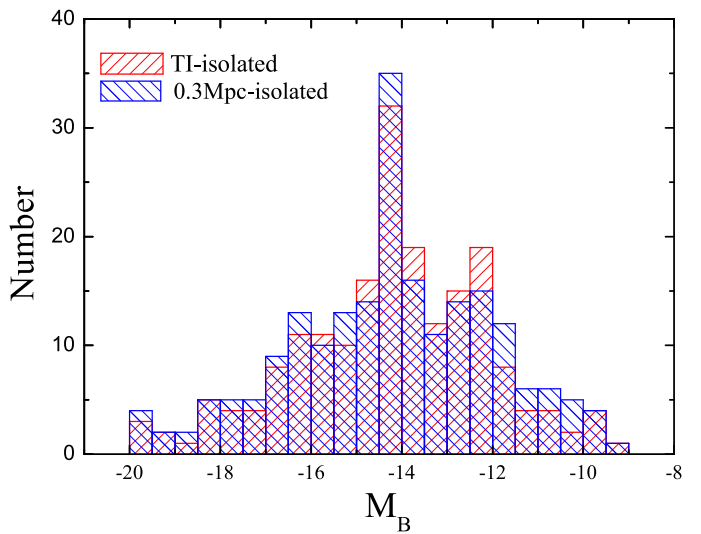}}
\caption{The number of isolated galaxies shown for bins of absolute
magnitude $M_B$. Either a geometrical neighbourhood criterion
($0.3$Mpc--isolated) or a dynamical criterion based on the tidal
index (TI) are used to determine isolated galaxies.
\label{fig:number_isolated}}
\end{figure}

This behaviour can be explained by looking at the luminosity function
of isolated galaxies in the samples.  Fig.\,\ref{fig:number_isolated}
shows the number of isolated galaxies in the local volume versus the
absolute B~magnitude.
We use two selection criteria for \emph{isolated galaxies}: first a
geometrical one -- an isolated galaxies may not have any neighbour
within a distance of 0.3\,Mpc, which is close to the virial radius of
an $M^*$ galaxy.
As a second criterion we use the tidal index (TI) of
\cite{karachentsev:catalog}, comparing a local dynamical timescale of
the galaxy to the Hubble time. We consider galaxies with a tidal index
less than zero as isolated.  As can be seen from
Fig.\,\ref{fig:number_isolated} the abundances of galaxies selected by
either geometrical or physical criteria match.  In both cases the same
prominent feature shows up, a peak in the number of isolated galaxies
with an absolute Magnitude $M_B\sim-14$.
This peak explains that mainly \emph{isolated} galaxies are lost in
going from the sample \texttt{lv8m14} with $M_B\le-14$ to the more
luminous sample \texttt{lv8m15} with $M_B\le-15$. It is reasonable to
assume that isolated galaxies predominantly reside in voids. Then the
voids are emptier in \texttt{lv8m15}, which is exactly what we observe
with the Minkowski functionals.

%%%
\subsection{Error estimates}
The position of a galaxy on the sky is known with a high
accuracy. However, the radial distance is estimated with some distance
indicator.  To quantify the influence of the distance error on our
estimates of the Minkowski functionals we randomise the radial
distance $r$ using a Gaussian probability law with mean $r$ and
standard deviation $0.1 r$.
We observe that the Minkowski functionals of the randomised samples
follow closely the functionals of the galaxy sample, well within the
error bars.
We compare this measurement error to the statistical error of
randomly distributed points with the same number density. This Poisson
error is approximately twice as large as the measurement error
obtained from randomising the distances.
Also the error from the subsampling (sometimes called Jackknife error,
see Fig.\,\ref{fig:minm_galm1415}) has approximately the same
amplitude as the Poisson error.

%%%%%%%%%%%%%%%%%%%%%%%%%%%%%%%%%%%
\section{Comparison with mock samples}
The local volume is certainly not a fair sample of the
Universe. However, only within our local neighbourhood we are able to
probe the galaxy distribution down to very low absolute luminosity.
In the local volume we have prominent structural elements like the
Tully void and the supergalactic plane. The extraction of mock samples
requires a careful selection of the position in the simulation box to
find comparable features.
Moreover the simulation has to be performed in a box large enough to
allow those structures to form and, on the other side, with enough
resolution on small scales to resolve the low mass halos.

%%%
\subsection{Simulation}
We use N-body simulation provided by A.~Klypin done with the Adaptive
Refinement Tree code \citep{kravtsov:adaptive}.  The simulation was
performed in a box with side lengths $160h^{-1}$\,Mpc for a spatially
flat cosmological $\Lambda$CDM model, which is compatible with the
3rd~year WMAP data \citep{spergel:wmap3} with the parameters $h=0.73$,
$\Omega_m=0.24$, $\Omega_\textrm{\tiny bar}=0.042$,
$\Omega_{\Lambda}=0.76$, and the normalisation $\sigma_8 = 0.75$.  In
this simulation the total number of particles is $1024^3$, the mass of
a particle is $3.18\cdot10^8h^{-1}M_{\odot}$, the spatial resolution
is $ 1.2h^{-1}$\,kpc, and the circular velocity of the smallest
resolved halo is 27~km/s.
We re-scaled all data (coordinates and masses of halos) to real units
assuming $H_0=72$km/s/Mpc, which is close to the value reported from
the WMAP analysis.
Halos are extracted using the bound density maximum halo finder
\citep{klypin:galaxies}, which detects bound halos with their subhalos
and calculates several physical and geometrical parameters of these
(sub)halos.
We use the maximum circular velocity $V_c$ to rank order the halos.
The maximum circular velocity $V_c$ can be determined more stable from
the simulations and is easier to relate to observations as compared
with the virial mass. We have a monotone relation between mass and
circular velocity $V_c$. For reference, halos with $V_c=50$\,km/s
have a virial mass of approx.\ $10^{10}M_{\odot}$ 
and halos with $V_c=20$\,km/s have a virial mass of approx.\
$10^{9}M_{\odot}$.

%*************************************************************
\subsection{Selection of a neighbourhood comparable to the local volume}
\label{sec:select-local}

The reasonably big volume of the simulation box allows us to select
mock samples that mimic the local volume features more closely.  As in
\cite{tikhonov:emptiness} we use several criteria to select spheres
with a radius of 8\,Mpc from the simulation box.
\begin{compactenum}
\item No halos with $\textit{mass} > 2\cdot10^{13}M_{\odot}$ reside
  inside the 8\,Mpc sphere.
\item The sphere must be centred on a halo with
  $1.5\cdot10^{12}M_{\odot} < \textit{mass} <
  3\cdot10^{12}M_{\odot}$\,km/s (the local group analog).
\item The number density of halos with $V_c>100$\,km/s inside the
  8\,Mpc sphere exceeds the mean value in the whole box by a factor 
  ranging from $1.5$ to $1.7$.
\item The number density of halos with $V_c>100$\,km/s found inside a
  4.5\,Mpc sphere exceeds the mean value in the whole box by a factor
  greater than 4;
\item There is no halo more massive than $5.0\cdot10^{11}M_{\odot}$
  with a distance in the range from 1 to 3\,Mpc).
\item The central halos of other mock samples are more distant than
16\,Mpc. There is no overlap between the samples.
\end{compactenum}
As discussed in \cite{tikhonov:emptiness} samples selected by the
above criteria may be considered as close geometrical and physical
cousins of the real local volume.
We could selected five such samples from the simulation box. Visual
inspection shows that our mock samples look quite similar to the local
volume galaxy sample.

%************************************************************
\subsection{Mapping between luminosity and circular velocity of the
halos}
\label{sec:mapping}
When assigning luminosities to dark matter halos we follow the
prescription of \citet{conroy:modeling}: First we build an ordered
list of all the galaxies in the observed local volume sample ranked by
their luminosity.  Then we rank order the halos in our local volume
candidate samples by their circular velocity.  Now the luminosity of
the brightest galaxy is assigned to the halo with the largest circular
velocity. Then the luminosity of the second brightest galaxy is
assigned to the second biggest halo and so on. This procedure
preserves the galaxy luminosity function and our mock samples have the
same number density as the observed galaxy sample (see
Table.~\ref{tab:gal}). By construction we have a monotonic relation
between $V_c$ and $M_B$. \citet{conroy:modeling} showed that this
prescription reproduces also the clustering properties of the more
massive galaxies in SDSS samples.
Nonetheless this mapping is not perfect. Only halos with
$V_c>30$\,km/s are left within the mock samples, whereas in the
observed sample galaxies with $M_B\approx -12$ may have a
significantly lower circular velocity \citep{tikhonov:emptiness}.
However if we would assign higher luminosities also to halos with
small circular velocities we would end up with too many halos.

%%%
\subsection{Morphological comparison}
\begin{figure*}
\sidecaption
\includegraphics[width=12cm]{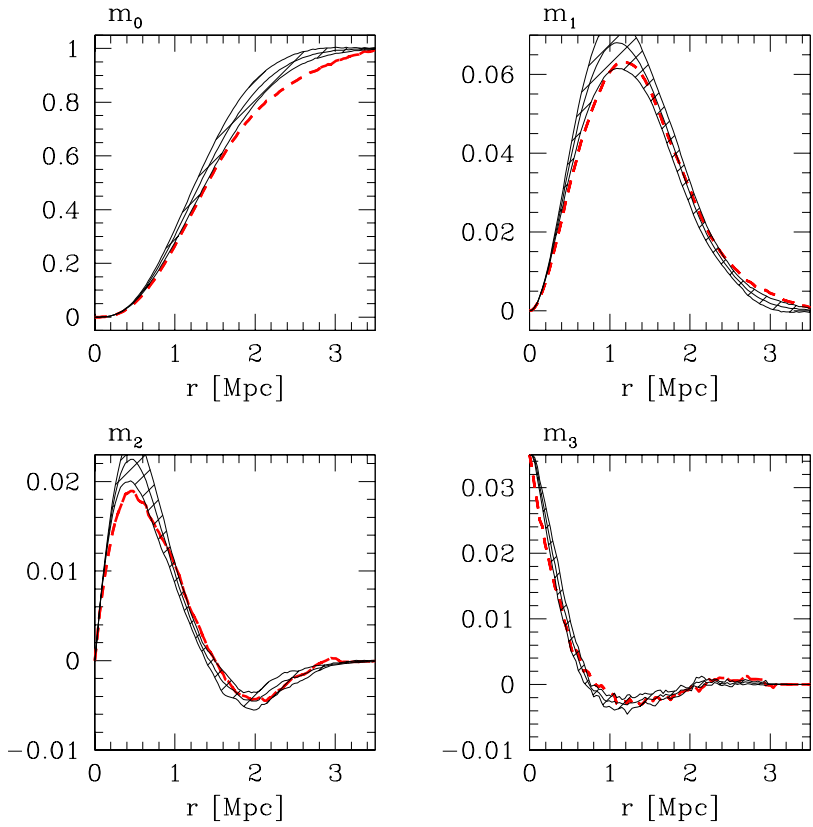}
\caption{Volume densities $m_\mu(A_r)$ of the Minkowski functionals
 determined from the galaxy samples with $M_B\le-12$ (short dashed red
 lines) compared to averages over the corresponding halo samples
 (shaded $1\sigma$ area).}
\label{fig:minm_galsimul_m12}
\end{figure*}

\begin{figure*}
\sidecaption
\includegraphics[width=12cm]{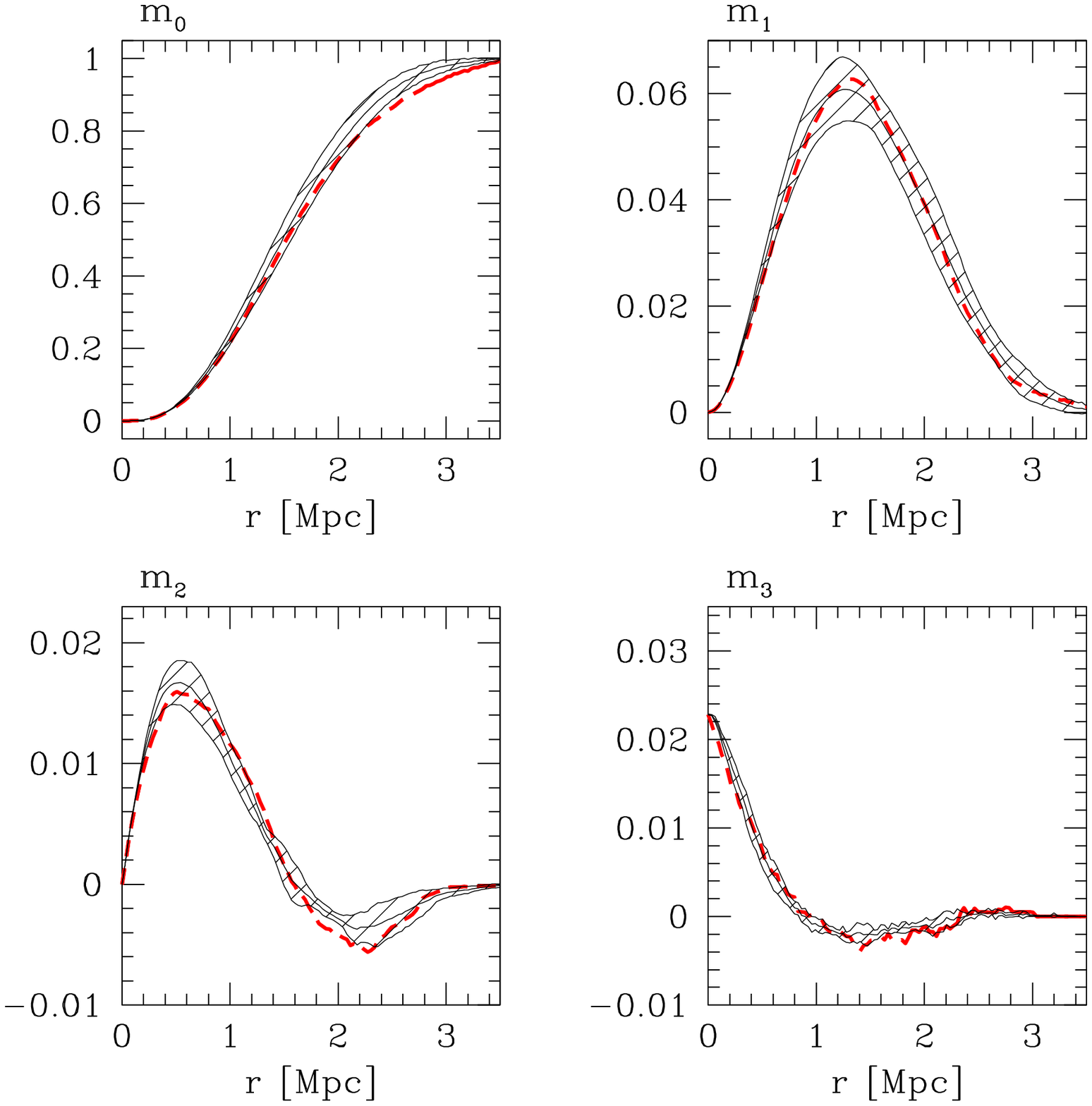}
\caption{Volume densities $m_\mu(A_r)$ of the Minkowski functionals
 determined from the galaxy samples with $M_B\le-14$ (short dashed red
 lines) compared to averages over the corresponding halo samples
 (shaded $1\sigma$ area).}
\label{fig:minm_galsimul_m14}
\end{figure*}

\begin{figure*}
\sidecaption
\includegraphics[width=12cm]{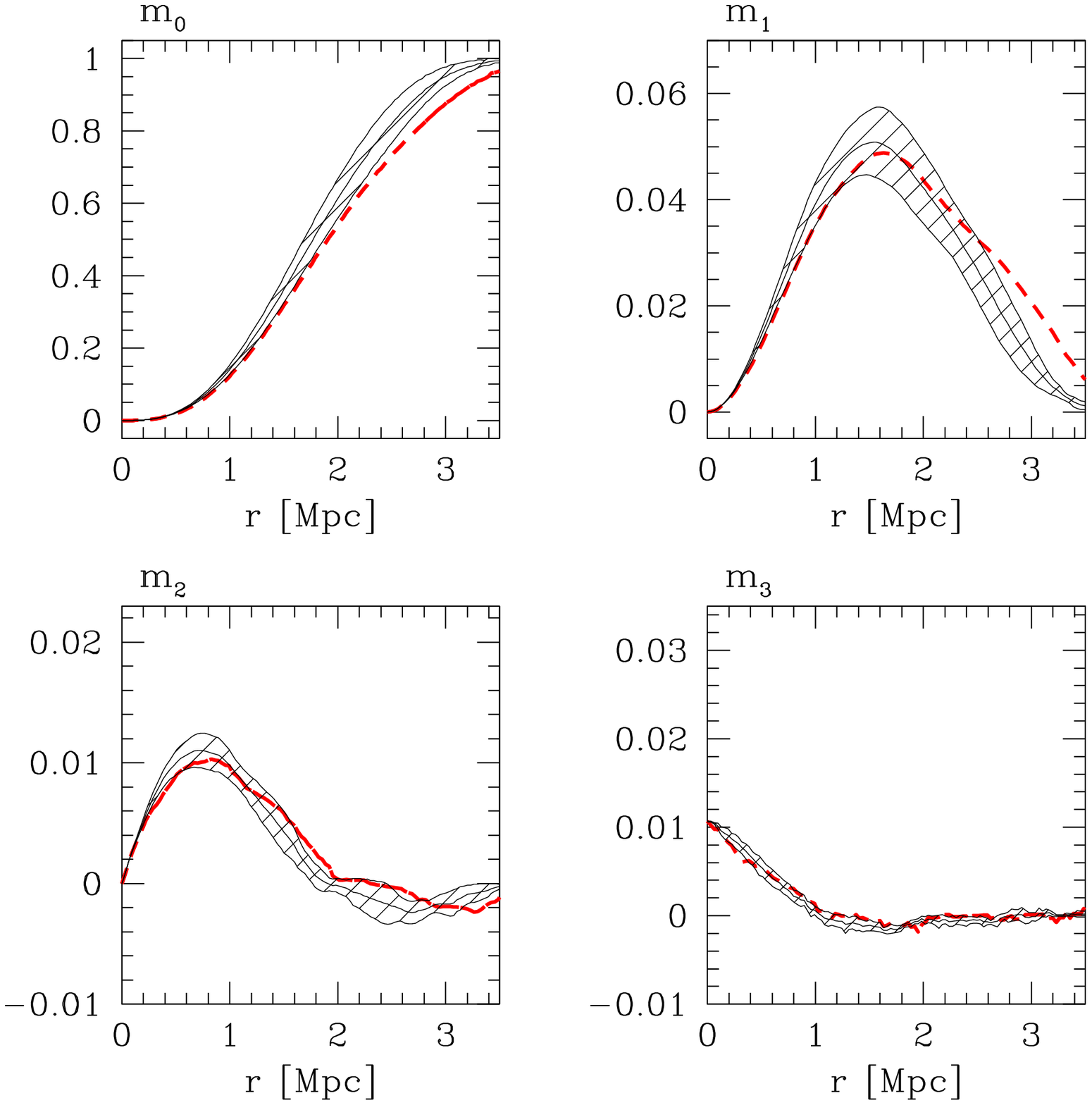}
\caption{Volume densities $m_\mu(A_r)$ of the Minkowski functionals
 determined from the galaxy samples with $M_B\le-16$ (short dashed red
 lines) compared to averages over the corresponding halo samples
 (shaded $1\sigma$ area).}
\label{fig:minm_galsimul_m16}
\end{figure*}

In Fig.\,\ref{fig:minm_galsimul_m12}--\ref{fig:minm_galsimul_m16} we
compare the morphology of the galaxy distribution to the morphology of
the corresponding halo samples from the $\Lambda$CDM simulation. We
show the average and the one--$\sigma$ range for the volume densities
of the Minkowski functionals estimated from the five mock samples.

The surface area, the integral mean curvature and the Euler
characteristic from the $M_B\le-14$ galaxy sample are well reproduced
by the corresponding mock samples as seen in
Fig.\,\ref{fig:minm_galsimul_m14}. Only the volume density $m_0$ is
slightly reduced on large scales, indicating emptier voids.

This changes if we compare with the samples including less luminous
galaxies with limiting absolute magnitude $M_B\le-12$ (see
Fig.\,\ref{fig:minm_galsimul_m12}).  The stronger clustering of the
galaxies is visible in the steeper decrease of the Euler
characteristic $m_3(A_r)$ for small $r$ and the reduced maximum of the
integral mean curvature $m_2(A_r)$. The well known fact, that the
voids are emptier in the real galaxy distribution, can be seen from
the reduced $m_0(A_r)$ for large $r$.

The more luminous halo samples with $M_B\le-16$ also shows some
morphological discrepancies compared to the corresponding galaxy
sample (Fig.\,\ref{fig:minm_galsimul_m16}).  $m_0(A_r)$ is reduced for
large radii, and we conclude that the voids are emptier in the real
galaxy distribution. Also for larger radii, the halo samples do not
show the pronounced coherent structures visible in the galaxy
distribution, as can be seen from the increased surface density
$m_1(A_r)$ of the galaxy samples as compared to the halo samples.

As we have seen in sect.~\ref{sec:morph-mag}, the morphology of the
observed galaxy distribution is depending on the limiting magnitude.
To investigate this behaviour in the mock halo samples we proceed
similarly to sect.~\ref{sec:morph-mag} and randomly subsample the halo
samples corresponding to $M_B\le-12$ and $M_B\le-14$ to the same
number density as seen in the $M_B\le-15$ sample.
The morphology of the halo distribution only shows a week trend, still
within the error bars, if we consistently select galaxies with higher
circular velocities. Hence, the physical mechanisms leading to these
emptier voids for $M_B\le-15$ in the observed galaxy distribution
(c.f.\ Fig.\,\ref{fig:minm_galm1415}) are not captured by the
$\Lambda$CDM model, and/or the mapping from the circular velocity to
the luminosity is more complicated at the faint end of the luminosity
function.

%%%%%%%%%%%%%%%%%%%%%%%%%%%%%%%%%%%
\section{Summary and Discussion}
To quantify the global morphology of the nearby galaxy distribution we
calculated Minkowski functionals from a series of volume--limited
samples extracted from the Local Volume galaxy catalogue (an updated
version of the Catalogue of Nearby Galaxies from
\cite{karachentsev:catalog}).  These samples are nearly complete for
$M_B\le-12$ within a distance of 8\,Mpc from our galaxy.
The strong clustering is dominating the morphology of the galaxy
samples on small scales. On larger scales the empty voids and coherent
structures show up in the Minkowski functionals.  It seems quite
obvious to attribute these morphological features to the Tully void,
covering approximately a quarter of the sample, and the supergalactic
plane.
By extracting a series of volume limited samples with increasing
luminosity we encountered a clear change in the global morphology of
the galaxy distribution. Going from a sample with $M_B\le-14$ to the
more luminous sample with $M_B\le-15$ the voids become significantly
emptier.
Indeed we saw a prominent peak in the luminosity function of isolated
galaxies with $M_B\approx-14$. We checked from the catalogue that all
these galaxies are dwarf irregulars with significant relative star
formation rates and hydrogen mass. Excluding these galaxies from a
sample makes voids significantly emptier.
It will be interesting to check if these galaxies form a unique
population and to which extend selection effects or true physical
effects cause the peak in the luminosity function.
We checked that the errors in the distance indicators do not influence
our results.

In the second part of the paper we compared the morphology of the
galaxy distribution to mock samples from a $\Lambda$CDM
simulation. The position of our mock samples inside the simulation box
was chosen to mimic the main features of our real galactic
neighbourhood, and the luminosities have been assigned to the halos in
such a way that the observational number density and luminosity
function are preserved.
The overall picture, with clustering on small scale, large
voids and coherent structures on large scales can be seen also in the
morphology of the mock samples. This agreement in principle 
allows us to look at the morphology in more detail.
In spite of the careful selection of the mock samples we still see
morphological differences between mock and galaxy samples. The
galaxies cluster stronger and the observed voids are emptier than in
the mock samples.
This observation is going beyond the well known overabundance of halos
in $\Lambda$CDM models. By construction we have the same number of
objects in the mock sample as in the galaxy sample, and we enforce the
same luminosity function. Hence we see differences in the geometry and
topology of the spatial distributions.
Clearly, five mock samples do not allow any elaborate statistical
tests (see e.g.\ \citealt{besag:simple}), but we may calculate a rough
estimate of the fluctuations in the Minkowski functionals. These
fluctuations are of the same order as observed for a Poisson process.
Therefore we claim to see differences in the morphology, only if the
Minkowski functionals are consistently outside the one--$\sigma$
range.

Certainly, our procedure of assigning luminosities to halos is
oversimplified.  Especially for low luminosities the mapping between
halo circular velocity and luminosity may not be straightforward. With
our mapping only halos with circular velocity $V_c > 30$\,km/s are
left in the mock samples. This is in agreement with theoretical
predictions for the least massive halo that can host a galaxy (see
e.g.\ \cite{hoeft:dwarf,loeb:light}).  However, in reality one
observes galaxies with $M_B$ in the range from {-12} to {-13} with
significantly lower rotational velocities \citep{tikhonov:emptiness}.
If we would assign higher luminosities to halos with small circular
velocities $V_c$ we either end up with too many mock galaxies, or we
would have to drop some halos with high $V_c$ to keep the number of
mock galaxies equal to the observed number. Such a procedure would
break the monotonic relation between $V_c$ and $M_B$.
To generate larger emptier voids in the mock samples, one would have
to additionally include halos with small $V_c$ residing in the
vicinity of filaments and sheets, and drop halos mainly residing in
voids. Also halos in voids with a small $V_c$ should remain dim.
A more intricate, environment dependent modelling of the luminosities
for the mock halos seems necessary. Especially if we also want to
understand the dependence of the global morphology on the absolute
magnitude cut in our galaxy samples.

Another factor explaining the morphological discrepancies might be that 
all the five mock samples we could extract do not hold a void which is
comparable in its size to the local (Tully) void.
This allows two lines of arguments, either one blames the observations
or the simulation.  Looking at Fig.\,\ref{fig:sky} we see that the
zone of avoidance (our galaxy) cuts right through the Tully void.
However, several targeted and blind searches of the zone of avoidance
have been performed. It is unlikely that we actually miss enough
galaxies to cut the void into halves \citep{karachentsev:catalog}.
On the other hand the simulation was conducted in a box with a side
length of $160h^{-1}$\,Mpc. This size is considered as reasonably
large, since we are only interested in mock samples within a sphere of
8\,Mpc radius. It remains an open question whether a $\Lambda$CDM
simulation in a larger box could provide us with mock samples with big
enough voids, especially when we consider the halo distribution at the
low mass end of the mass function.

%%%%%%%%%%%%%%%%%%%%%%%%%%%%%%%%%%%
\begin{acknowledgements}
  We thank I.D.\,Karachentsev for providing us an updated list of his
  catalogue of neighbouring galaxies and A. Klypin for providing
  results of computer simulations that were conducted on the Columbia
  supercomputer at the NASA Advanced Supercomputing Division. 
  Anton Tikhonov acknowledges support from the Deutsche
  Forschungsgemeinschaft (DFG grant: GO 563/17-1), and the ASTROSIM
  network of the European Science Foundation (ESF) and wishes to thank
  the Astrophysical Institute Potsdam for their hospitality.
\end{acknowledgements}

%%%%%%%%%%%%%%%%%%%%%%%%%%%%%%%%%%%
\appendix
\section{Calculating Minkowski functionals}
\label{sec:calculating}

The code used for the calculations of the Minkowski functionals is an
updated version of the code developed by \cite{kerscher:abell}, based
on the methods outlined in \cite{mecke:robust}. The code is made
available to the general public via
http:$//$www.math.lmu.de$/\!\sim$kerscher$/$software$/$

In the following we present the boundary corrections
used by \cite{schmalzing:minkowski}.
All galaxies in our samples are within a sphere of radius $R=8$\,Mpc
around our galaxy, and similar for the halo samples. Let $D_R$ be this
spherical window containing $N$ galaxies.  $A_r=\bigcup_{i=1}^NB_r(i)$
is the union of balls $B_r(i)$ of radius $r$ centred on the $i$--th
galaxy, respectively.

\begin{figure}\begin{center}
\resizebox{0.8\hsize}{!}{\includegraphics{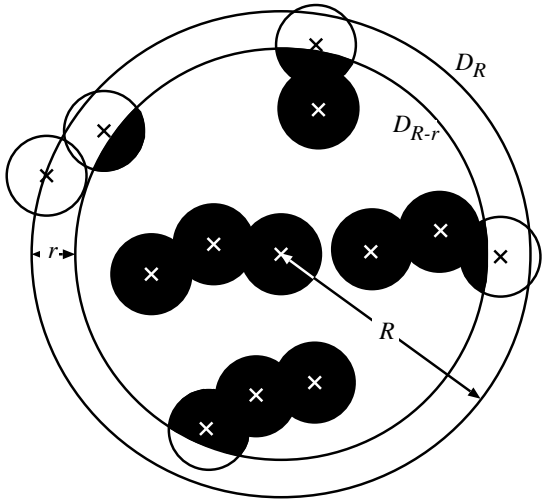}}
\caption{A two--dimensional sketch illustrating the geometry of the
boundary corrections used. $R=8$\,Mpc is the radius of the sphere
$D_R$ centred on our galaxy enclosing the local sample.  $r$ is the
radius of the individual balls centred on each galaxy, and the black
area is the set $A_r\cap D_{R-r}$.}
\label{fig:boundary}
\end{center}\end{figure}

Considering only $A_r$ we miss contributions from galaxies outside
$D_R$, i.e.\ from galaxies not included in our sample. A well defined
way is offered by looking at the intersection $A_r\cap D_{R-r}$
\citep{mecke:euler}. Using the shrunken window $D_{R-r}$ we make sure
that all possible contributions are taken into account, as illustrated
in Figure~\ref{fig:boundary}.  In order to measure the boundary
contributions we have to calculate the Minkowski functionals
$M_\mu(A_r\cap{D_{R-r}})$ of the intersection of the union of all
balls with the window, and the Minkowski functionals
$M_\mu(D_{R-r})$ of the window $D_{R-r}$ itself,

Now imagine the union set $\widetilde A_r$ of balls on all galaxies
in a very large (but still finite) patch of the Universe, with
$A_r\subset\widetilde A_r$.  Integrating over all movements
(translations and rotations) of the window within this large patch of
the universe we obtain the spatial average
\begin{equation}
\left\langle M_\mu\left(\widetilde A_r \cap D_{R-r}\right) \right\rangle =
\int_{\mathcal{G}} M_\mu\left(\widetilde A_r \cap gD_{R-r}\right) {\rm d} g.
\end{equation}
Here $gD_{R-r}$ is a translation and/or rotation of the shrunken window,
$\mathcal{G}$ is the group of motions, and ${\rm d}g$ the suitably
normalised Haar measure on this group.
The integral over motions can be expressed as a sum using the
principal kinematical formula \citep{hadwiger:vorlesung}
\begin{equation}\label{eq:kinematical}
\int_\mathcal{G} M_\mu\left(\widetilde A_r\cap gD_{R-r}\right) {\rm d}g =
\sum_{\nu=0}^\mu \binko{\mu}{\nu} M_\nu(\widetilde A_r) M_{\mu-\nu}(D_{R-r}).
\end{equation}
For a homogeneous and isotropic galaxy distribution the spatial
average coincides with the expectation over different realizations, and
$M_\mu(A_r\cap{D_{R-r}})$ is already an unbiased estimate of
$\left\langle M_\mu\left(\widetilde A_r \cap D_{R-r}\right)
\right\rangle$.
This observation together with eq.~(\ref{eq:kinematical}) allows us to
remove the boundary contribution of the window. For the volume density
this is almost trivial:
\begin{equation}
m_0(A_r) = \frac{M_{0}(A_r\cap D_{R-r})}{M_0(D_{R-r})}.
\end{equation}
For all the Minkowski functionals estimates for the volume densities
can be obtained by applying the following recursive
formula\footnote{we use the convention $\sum_{n=i}^j x_n = 0$ for
$j<i$} \citep{fava:plate}:
\begin{equation}\label{eq:min-deconv}
m_\mu(A_r) =
\frac{M_\mu(A_r \cap D_{R-r})}{M_0(D_{R-r})} -
\sum_{\nu=0}^{\mu-1} \binko{\mu}{\nu}
m_\nu(A_r) \frac{M_{\mu-\nu}(D_{R-r})}{M_0(D_{R-r})}.
\end{equation}
This linear transformation of the observed $M_\mu(A_r \cap D_{R-r})$
facilitates the direct comparison with the volume densities
$m_\mu\left(\widetilde A_r\right)$ calculated from Point process
models, namely the Poisson process.

%%%%%%%%%%%%
\section{Morphology with Minkowski functionals}
\label{sec:models}

\begin{figure}\begin{center}
\resizebox{\hsize}{!}{\includegraphics{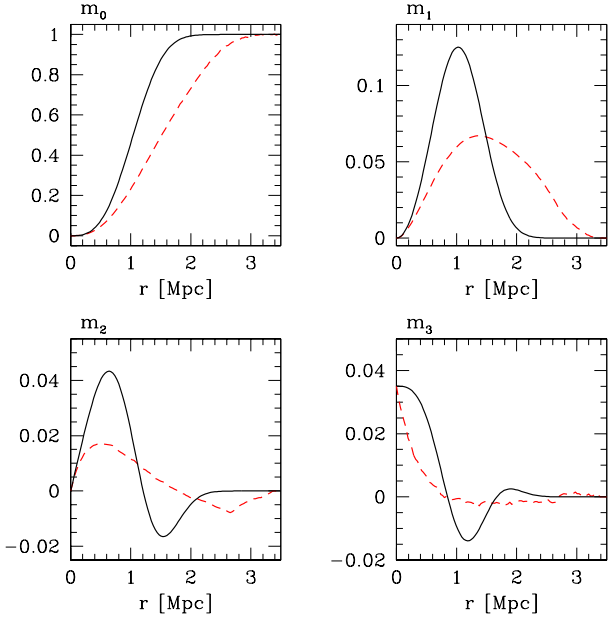}}
\caption{The volume densities $m_\mu(A_r)$ of the Minkowski
  functionals determined from a Poisson process (solid black lines)
  and points randomly distributed on random line segments (dashed red
  lines).  Both models have the same number of points, and the same
  sample geometry as the galaxies in \texttt{lv8m12}.}
\label{fig:min_line}
\end{center}\end{figure}

In the following we give a qualitative discussion of the Minkowski
functionals for a Poisson process decorated with balls of radius $r$
and relate them to the observed morphological features. The Minkowski
functionals of randomly distributed points (a Poisson process) are
shown in Fig.\,\ref{fig:min_line}.

With increasing radius $r$ the volume is filled until reaching
complete coverage where the volume density $m_0(A_r)$ reaches unity.
The surface density $m_1(A_r)\times8$ increases with the radius $r$,
reaches a maximum, and finally approaches zero when all the volume is
filled up and no surface area is left.  
For small radii the balls grow outwards and the positive density of
the integral mean curvature $m_2(A_r)2\pi^2>0$ is a reflection of
the mainly convex structure of the union set of balls. The volume
starts to fill and the integral mean curvature reaches a maximum. For
even larger radii we get negative contributions to the density of
integral mean curvature stemming from concave structures -- the balls
are growing inwards into the voids.  Finally $m_2(A_r)$ approaches
zero when no surface area is left.  
For the radius $r=0$ each ball (point) gives a contribution of one to
the Euler characteristic and the density of the Euler characteristic
$m_3(A_0)\frac{4\pi}{3}$ equals the number density of the sample.
With increasing radius more an more balls overlap and the Euler
characteristic decreases. Then tunnels through the structure are
forming, giving a negative contributions to $m_3(A_r)$.
As a clear signal of completely enclosed voids in the Poisson process,
$m_3(A_r)$ shows a second positive maximum.

Beyond the completely random model one may consider geometrically
inspired models, e.g.\ points randomly distributed on randomly
placed line segments. Minkowski functionals for such a low
dimensional Poisson processes embedded in three dimensions are
discussed by \cite{schmalzing:cfa2}.
As an example we show the Minkowski functionals for a model using in
the mean 2.1 points, randomly distributed on randomly placed line
segments with a length of 8\,Mpc. As can be seen from
Fig.~\ref{fig:min_line} some of the morphological features of the
galaxy distribution discussed in Section~\ref{sec:morph-local-vol} can
be reproduced.  Specifically the stronger clustering and the
suppression of completely enclosed voids is clearly seen in the
Minkowski functionals. The choice of these parameters does not seem
realistic. Choosing more points per line segment or using shorter
segments leads to a largely different values for the Minkowski
functionals.
Indeed \cite{schmalzing:cfa2} have shown that only with (random)
mixtures of similar Poisson models for filament--, sheet-- and
field--galaxies one can reproduce the small scale morphology of the
galaxy distribution. As can be seen from their Fig.\,B1 these mixture
models are not appropriate on larger scales. This is expected, since
randomly placed geometrical objects will not be able to model coherent
large scale features.
Other models are based on the hierarchy of correlation functions.
The explicit incorporation of higher moments for the construction of
point process models is discussed in \cite{kerscher:constructing}. At
least for the lowest order models the Minkowski functionals can be
given explicitly \citep{kerscher:reflex}.
Models, motivated from statistical physics and spatial
statistics are discussed by \cite{mecke:additivity} and
\cite{mecke:morphological}.

\end{document}